\documentclass[nofootinbib,preprintnumbers,12pt]{revtex4}

\usepackage{amsmath,amsfonts,amsthm,amssymb,epsf}
\usepackage{dcolumn}
\usepackage{graphicx}
\usepackage{wasysym}
\usepackage{subfigure}
\usepackage{multirow}

\newcommand{\beqa}{\begin{eqnarray}}
\newcommand{\eeqa}{\end{eqnarray}}

\newcommand{\ra}{\rightarrow}
\newcommand{\da}{\dagger}

\newcommand{\cln}{\colon}

 \def\HH{{\cal H}}

\def\be{\begin{equation}} \def\ee{\end{equation}}

\begin{document}

\title{Quantum Graphity:  a  model of emergent locality}

\author{Tomasz Konopka${}^1$, Fotini Markopoulou${}^{2,3}$ and Simone Severini${}^{3}$}

\affiliation{${}^1$ ITP, Utrecht University,
Utrecht 3584 CE, The Netherlands,\\ and \\ ${}^2$ Perimeter Institute for Theoretical Physics, \\
Waterloo, Ontario N2L 2Y5
Canada, \\
and \\ ${}^3$ University of Waterloo, Waterloo, Ontario N2L 3G1,
Canada.}

\preprint{ITP-UU-08/02}\preprint{SPIN-08/02}

\begin{abstract}

Quantum graphity is a background-independent model for emergent
macroscopic locality, spatial geometry and matter. The states of
the system correspond to dynamical graphs on $N$ vertices. At high
energy, the graph describing the system is highly connected and
the physics is invariant under the full symmetric group acting on
the vertices. We present evidence that the model also has a
low-energy phase in which the graph describing the system breaks
permutation symmetry and appears to be ordered, low-dimensional
and local. Consideration of the free energy associated with the
dominant terms in the dynamics shows that this low-energy state is
thermodynamically stable under local perturbations. The model can
also give rise to an emergent $U(1)$ gauge theory in the ground
state by the string-net condensation mechanism of Levin and Wen.
We also reformulate the model in graph-theoretic terms and compare
its dynamics to some common graph processes.

\end{abstract} \vfill

\maketitle

\section{Introduction}

It is possible that the successful quantum theory of gravity will
require a modification of general relativity or quantum theory and
that at least one of the two is not fundamental but rather only an
effective, emergent theory.  Almost all approaches to quantum
gravity leave quantum theory intact and the suspicion is largely
on general relativity being the effective theory.
Establishing this is, however, a major challenge. General
relativity describes gravitation as the curvature of spacetime by
energy and matter, which means that, if it is only an effective
theory, then spacetime must be just an effective description of
something more fundamental. The trouble with this is that most of
known physics is formulated in terms of a spacetime geometry.

In approaches where general relativity is considered fundamental
enough to hope to obtain a quantum theory of gravity by its
quantization (such as causal dynamical triangulations \cite{CDT},
loop quantum gravity \cite{LQG}, spin foam models \cite{SF} and
group field theory \cite{OritiGFT}), one needs a mechanism to
generate a nearly-flat, classical geometry in the low-energy
limit, complete with local observables, to compare theory with
experiment. While progress has been made in such approaches, there
are still open questions. In approaches with extra dimensions, one
would like an explanation for why some dimensions are large and
others small \cite{KalyanaRama:2006xg}. In the realm of emergent
gravity approaches, we encounter theories that are formulated in
terms of quantum fields on a given geometry (this includes
condensed-matter and analog approaches \cite{Vol,Vis,Unr,CalHu},
matrix models \cite{Ban,Hor} and certain formulations of string
theory \cite{Mal,GKP}). The evidence for emergent gravity is, for
example, in the form of a spin-2 field, an effective metric, or
the anti-de Sitter/conformal field theory (AdS/CFT) duality
\cite{Mal}. Inspecting these approaches, however, we find that it
is unclear to what extent the geometry used in the initial
formulation and its symmetries enter the results. Is the initial
fixed geometry an auxiliary structure or does it have a physical
meaning?

A related issue is the notion of locality in a quantum theory of
gravity. Locality is a universal property of known physics so it
is natural that we have also been looking for a local quantum
theory of gravity. However, there are a number of indications that
this may not be correct (a thorough investigation of this question
can be found in \cite{Gid}). In addition, some of the main
obstacles we encounter in approaches to quantum gravity can be
traced to the problem of constructing local observables that
quantum gravity inherits from general relativity:  there are no
local diffeomorphism-invariant observables for pure gravity
\cite{Gid}. This problem has become more prominent in recent years
because its resolution is necessary to compare theory to
experiment.  We believe the question of locality should be
addressed by emergent gravity approaches. One may be justified to
expect that, if  gravity and geometry are emergent, so must be
locality.

A condensed-matter approach to the problem of emergent geometry
has recently been proposed through a model called {\em quantum
graphity} \cite{graphity1}. In that model, states of the system
are supported on the complete graph $K_N$ on $N\gg 1$ vertices in
which every two vertices are connected by an edge. Quantum degrees
of freedom are associated with edges of the graph: there is a
state for each edge which signifies that the edge is turned off
and other states which indicate that the edge is on. Thus, the
states of the complete system include every possible graph on $N$
vertices. The model describes a {\em dynamical graph} as the
answer to the question of whether two vertices in the graph are
adjacent or not can vary in time.

At high energy, there is no notion of geometry, dimension or
topology in the system. At low energy, however, the system is
expected to become ordered in such a way that the subgraph of
$K_N$ consisting of the ``on'' edges can be described in terms of
a low-dimensional spacetime manifold. Near this ground state, the
model is closely related to the string-net model of Levin and Wen
\cite{LevinWen} which has emergent $U(1)$ gauge degrees of freedom
coupled to massive charge particles. The transition process from
high to low temperature, called {\em geometrogenesis}, establishes
an emergent notion of locality in the low-energy regime. It is
worth emphasizing that the model is not ``nonlocal'' in the sense
of adding nonlocal corrections to a local theory.

In this paper we present a slightly modified and simplified
version of the quantum graphity model. Compared to the original
model, the version in this paper has a reduced state space
associated with each edge. This allows us to better concentrate on
the structural properties of the graph at low energies. The
dynamics of the model is also somewhat different from the original
so that there is more natural accommodation of features of the
graph such as counting of closed paths. For a certain set of
parameters, we present evidence that a graph with hexagonal
symmetry is at least a local free-energy minimum for the model.
The very interesting question of whether the system can generate a
three-dimensional lattice in some region of its parameter space is
left for future work.

It is useful to relate quantum graphity to existing approaches to
quantum gravity. It is certainly the case that several of the
so-called background-independent approaches to quantum gravity are
graph-based: Loop Quantum Gravity\cite{LQG}, Causal
Sets\cite{CauSet}, Algebraic Loop Quantum
Gravity\cite{Giesel:2006uj}, and Quantum Causal
Histories\cite{QCHistories}, among others. This is not surprising,
since network-based states have a strong relational character, a
feature considered desirable in a background-independent context.
Quantum graphity also shares with these theories a common central
question, the search for the semiclassical, or low-energy, states
in the theory. However, there are also basic differences. The
dynamics of quantum graphity with matter is essentially an
extension of the string-net Hamiltonian of Levin and Wen and not a
quantization of the Einstein equations (string-nets are tensor
product categories, just like spin networks, making the
introduction of Levin-Wen-type dynamics technically
straightforward). Additionally, the data on the network do not
correspond to SU(2) labels found on spin network states in Loop
Quantum Gravity. In quantum graphity, geometry is identified at
the low-energy phase from properties of the network itself.

The outline of the article is as follows. In Sec. \ref{s_model},
we define the model by putting states $|0\rangle, |1\rangle$ on
the edges of the complete graph $K_N$. A $|0\rangle$ state means
the edge is ``off,'' or missing, and a $|1\rangle$ state means the
edge is on. In Sec. \ref{s_q_hamiltonian}, we give the Hamiltonian
of our model. In Sec. \ref{s_largegraph}, we study the model when
the number of vertices $N$ is large and find that the hexagonal
lattice is a good candidate for the ground state for an
appropriate choice of parameters. We consider perturbations over
the ground state and find that the hexagonal lattice is
thermodynamically stable under local perturbations. In Sec.
\ref{s_moredof}, we introduce a degeneracy of the on edges:  the
$|1\rangle$ state is split into
$|1,0\rangle,|1,-1\rangle,|1,+1\rangle$.  This allows us to
introduce the string-net condensation mechanism of Levin and Wen
\cite{LevinWen} into our dynamics, bringing the model closer to
the original quantum graphity system \cite{graphity1}. In Sec.
\ref{s_graphtheory}, we initiate a reformulation of our model in
graph-theoretic terms, and provide some first observations on the
transition from the high- to the low-energy phase. In particular,
we compare the transition with processes generating random graphs.

Our model introduces a novel mechanism for emergent space and
locality and this comes with a new set of questions that need to
be investigated in future work. These include the role of time,
temperature, the actual transition between the two phases, and its
remnants. We discuss these in the concluding Sec.
\ref{s_discussion}.

\section{Graph Models\label{s_model}}

Graph-based, instead of metric-based, theories are attractive
implementations of the relational content of diffeomorphism
invariance. The interpretation is that it is the structure of the
graph, i.e. the relations between the constituents, that is
important to describe physics. As such, graphs are probably the
most common objects that appear in background-independent theories
of quantum gravity \cite{CDT,LQG,SF,CauSet,Llo}. Furthermore, it
has been previously argued in the literature that at the discrete
level, spacetime diffeomorphisms should appear as permutation
invariance of these fundamental constituents \cite{Sta}. We shall
implement this by starting with the complete graph $K_N$ on $N$
vertices, an object that is permutation invariant. The dynamics on
the complete graph will be chosen so that it respects the
permutation invariance of $K_N$ and depends on natural graph
features: vertices, closed paths and open paths.

We first review some useful graph-theoretic properties and
techniques. Next, we introduce the necessary quantum mechanical
notation and then finally define the Hamiltonian models on graphs
that we will consider in this paper.

\subsection{Graph theory preliminaries \label{s_graph_prelim}}

The complete undirected graph on $N$ vertices is denoted by $K_N$.
It is a graph in which every two vertices are connected by an
edge. If the vertices are labeled by $1,2,\ldots,N,$ then $K_N$
has an edge $e_{ab}$ connecting any two $a$ and $b$.

Any graph $G$ on $N$ vertices can be regarded as a subgraph of the
complete graph $K_N$; specifically, it can be obtained by deleting
edges from $K_N$. A convenient way to represent $G$ is via its set
of edges $E(G)$ or via its $N\times N$ adjacency matrix\be
A_{ab}(G)=\left\{ \begin{array}{ll} 1&\qquad{\mbox{if } }e_{ab}\in E(G)\\
0&\qquad {\mbox{otherwise.}} \end{array} \right. \ee
By definition, the adjacency matrix is symmetric and it has zero diagonal.

Information about a graph can be obtained from its adjacency
matrix with the use of linear algebra. In particular, powers of
the adjacency matrix, defined as follows: \be \label{Apowers}
\begin{split} A^{(2)}_{ab} &= \sum_c \, A_{ac} A_{cb}, \\
A^{(3)}_{ab} &= \sum_{c}\sum_d \, A_{ac} A_{cd} A_{db}, \;\;\;
\mathrm{etc.},
\end{split} \ee contain information about open and closed paths in
the graph. As an example, the $ab$ component of the $n^{th}$ power of
$A$ denotes the number of ways one can move from vertex $a$ to
vertex $b$ by jumping only along the edges of the graph in a fixed,
$n$, number of steps. When $a=b$ and the element considered is on
the diagonal, then $A^{(n)}_{aa}$ denotes the number of paths in the
graph of length $n$ that start and end on the same vertex $a$.

When using the powers of the adjacency matrix to enumerate closed
and open paths, it is essential to understand that the numbers
computed include paths which traverse certain edges more than
once. For example, in a graph with vertices labeled by $1,2,...,N$
and edges labeled by pairs $\{\{1,2\},\{1,3\},\{2,3\},...\}$, a
sequence such as $\{\{1,2\}, \{2,1\}, \{1,2\}, \{2,3\}\}$ would be
counted as a path of length four from vertex $1$ to vertex $3$,
irrespective of the fact that the edge $\{1,2\}$ is used $3$ times
in the sequence or that there exist a shorter sequence of edges,
namely $\{\{1,2\}, \{2,3\}\}$, that connects the same two
vertices.

For future use, it is also useful to define the notion of
nonretracing paths. We define a \emph{nonretracing path} to be an
alternating sequence of vertices and edges, in which any
particular edge appears exactly once. It is useful to specify that
nonretracing paths can be \emph{open} or \emph{closed} and that a
nonretracing path is not necessarily a geodesic between two
vertices. A closed nonretracing path is also said to be a
\emph{cycle}. The number of cycles can be computed algorithmically
but not with the straight forward use of powers of the adjacency
matrix. Some questions regarding counting the number of cycles of
a given length can be very difficult (see, \emph{e.g.},
\cite{GJ}).

\subsection{Quantum mechanics preliminaries \label{s_qm_prelim}}

We would now like to set up a framework which would allow us to
encode the complete graph $K_N$ and its subgraphs as states in a
quantum mechanical Hilbert space. To do this, we define a large
Hilbert space $\HH_{total}$ made up of smaller spaces associated
with components of the complete graph $K_N$. In general, it is
possible to associate a Hilbert space $\HH_{edge}$ to each edge
$e_{ab}$ and a Hilbert space $\HH_{vertex}$ to each vertex. The
total Hilbert space of the system would then be the tensor product
\be \HH_{total} = \label{HHtot} \bigotimes^{N(N-1)/2} \!\!\!\!
\HH_{edge} \;\, \bigotimes^{N} \; \HH_{vertex}. \ee In the
following, we specialize to models in which all the degrees of
freedom are on the edges of the graph as opposed to both the edges
and vertices.

The basic Hilbert space associated with an edge is chosen to be
that of a fermionic oscillator. That is, $\HH_{edge}$ will be \be
\label{HHedge} \HH_{edge} = \mathrm{span}\{ |\,0\rangle, \,
|\,1\rangle \}; \ee the state $|\, 0\rangle$ is called the empty
state and the state $|\, 1\rangle$ is said to contain one
particle. (One can alternatively think of $|\,0\rangle$ and $|\,
1\rangle$ as being states in the computational basis of a qubit.)
A general state in the total space of edges $\HH_{edge}^{\otimes
N(N-1)/2}$ is \be \label{psiGgeneral} |\psi\rangle = \sum_{\{n\}}
\, c_{\{n\}} \,|n_{12}\rangle \otimes |n_{13}\rangle \otimes
|n_{23}\rangle \otimes \cdots, \ee \emph{i.e.}, a superposition of
all possible states which are themselves tensor products of states
$|n_{ab}\rangle$ associated with single edges; $n_{ab}=0,1$ are
occupation numbers and $c_{\{n\}}$ are complex coefficients.

In the graph model, a given edge of the graph is interpreted as
being on or off depending on whether the corresponding state has a
particle or not. The collection of on states define a subgraph of
the complete graph $K_N$. Thus, the total Hilbert space of edges
can be decomposed as (recall that we ignore degrees of freedom on
the vertices) \be \label{HHtotsum} \HH_{total} = \bigoplus_{G}
\HH_{G} \ee with the tensor sum being over all subgraphs $G$ of
$K_N$. Each term in (\ref{psiGgeneral}) corresponds to a state in
one of the blocks $\HH_G$. Since we treat the vertices as
distinguishable, there may be many blocks in the sum that
correspond to isomorphic graphs.

Acting on the Hilbert space of each edge are the usual creation
and annihilation operators $a^\dagger$ and $a$. They act in the
usual way, \be a|\,0\rangle = 0, \qquad a|\,1\rangle = |0\rangle,
\ee and obey the anticommutation relation \be \{a, a^\dagger\} =
aa^\da + a^\da a = 1.\ee The other anticommutators are zero, $\{
a,a\}=\{a^\da, a^\da\}=0 $. There is a Hermitian operator $a^\da
a$, whose action on a state $|n\rangle$ with $n=0,1$ is \be
\label{NumberOp} a^\da a |\,n\rangle = n\, |\, n\rangle. \ee This
operator is commonly called the number operator because it reveals
the number of particles present in a state.

It is now possible to define operators (\ref{NumberOp}) that act on
each of the copies of $\HH_{edge}$. These will be denoted by
subscripts and defined in the intuitive way, \emph{e.g.} \be
\begin{split} N_{13} \, & \left(|n_{12}\rangle \otimes
|n_{13}\rangle \otimes \cdots\right) \\ &\;\;\, = (1\otimes a^\da
a \otimes \cdots)\, \left(|n_{12}\rangle \otimes
|n_{13}\rangle \otimes \cdots\right) \\
&\;\;\,= n_{13} \, \left(|n_{12}\rangle \otimes |n_{13}\rangle
\otimes \cdots\right). \end{split}\ee From the definition of the
operators on the middle line, one can see that number operators
acting on different edges commute. Also, since the graphs we are
considering are undirected (that is, the edges are unordered pairs
of vertices), we identify $N_{ab} = N_{ba}$.

Note that the set of operators $N_{ab}$ can be understood as
analogous to elements of an adjacency matrix $A_{ab}$. That is,
the operator $N_{ab}$ gives zero when the state of edge $e_{ab}$
is off and gives $1$ when that edge is in an on state. In the
previous section it was shown that it is very useful to define
powers of the adjacency matrix as in (\ref{Apowers}). It is also
reasonable to introduce powers of these number operators. For
example, we define \be \label{Npowers}
\begin{split} N^{(2)}_{ab} &= \sum_c \, N_{ac} N_{cb}, \\ N^{(3)}_{ab} &=
\sum_c\sum_d \, N_{ac} N_{cd} N_{db}, \;\;\; \mathrm{etc.}
\end{split} \ee When the elements of these matrices $N_{ab}^{(L)}$ act on a
state, they return a nonzero answer if the state contained a path
between two vertices of a certain length $L$ passing through edges
whose $n$ values is different from zero. Thus these operators are
quantum mechanical analogs of $A_{ab}^{(L)}$ that count the number
of closed and open paths that pass through a vertex; here these
operators count closed and open paths in the on graph only.

There are some differences, however, due to the fact that the
creation and annihilation operators $a_{ab}$ and $a_{ab}^\da$
acting on the same edge do not commute. Terms which contain at
least two creation operators and two annihilation operators can in
principle be ordered in several inequivalent ways. In setups
involving the harmonic oscillators, there is a standard convention
for ordering operators called {\em normal ordering} denoted by
putting colons around an operator. In this convention, all
annihilation operators $a_{ab}$ are set to the right of the
creation operators $a^\da_{ab}.$ For example, the terms in the
normal-ordered number operator squared are of the form \be \cln
N_{bc} N_{cd} \cln = \cln a^\da_{bc} a_{bc} a^\da_{cd} a_{cd} \cln
= a^\da_{bc} a^\da_{cd} a_{bc} a_{cd}. \ee When $b=d$, the same
two annihilation operators appear on the right. Since (for
$n=0,1$) \be a a |\,n\rangle = 0, \ee which follows from the
anticommutation relations, one finds that \be
\label{Numbersquared} \cln N_{bc} N_{cb} \cln = 0. \ee
Consequently, whenever a term of $\cln N_{ab}^{(L)} \cln$ with
$L\geq 2$ acts on the same edge more than once, that term does not
contribute. Therefore, the eigenvalues of operators $\cln
N_{ab}^{(L)}\cln$ for each $a,b$ return the number of {\em
nonoverlapping paths} between vertices $a$ and $b$. We will make
use of the normal ordering convention and this property, in
particular, when defining and analyzing the quantum Hamiltonian
for the graph model.

\subsection{Hamiltonian \label{s_q_hamiltonian}}

We would now like to define a condensed matter like model in which
the configuration space is the space of all possible graphs on a
fixed number of vertices.

For this purpose we consider Hamiltonian function (operator) $H$
acting on states in the Hilbert space $\HH_{total}$ defined in
(\ref{HHtot}). A Hamiltonian operator is usually used to associate
an energy $E(G)$ with a state $|\,\psi_G\rangle$. We do this here
using the normal-ordering prescription described above, \be
\label{EGnormal} E(G) = \langle \psi_G | \, \cln H\cln |\,
\psi_G\rangle . \ee This notation for the energy should not be
confused with the set of edges of a graph; the meaning of the
symbol $E(G)$ should be clear from the context.

We would like the Hamiltonian to preserve the permutation
invariance symmetry of $K_N$. In a general manner, therefore, we
can ask what Hamiltonian can be written down for a graph model
using the adjacency matrix operators defined in the previous
section. It turns out that there are many terms that can be
written down that fit the requirement of permutation symmetry. The
trace $\sum_{a} N_{aa}$ or the sum of the off-diagonal elements
$\sum_{a, b\neq a} N_{ab}$ are simple examples. Tracing or summing
over all elements can also be done using powers of $N_{ab}$, which
can be defined as in (\ref{Npowers}). Other possibilities include
first defining an object $N_a = \sum_b N_{ab}$, taking powers of
this object as in $N_a^{(p)} = (N_a)^{(p)}$ for some $p$, and then
summing $\sum_a N_a^{(p)}$. In the following we choose terms that
appear natural from the graph-theoretic perspective.

\subsubsection{Valence term}

A basic property of a graph is its distribution of vertex degrees
- the number of edges adjoining each vertex. Indeed, in graph
theory one often studies the class of $d$-regular graphs in which
all vertices have a specified and fixed degree (also called
valence). We would like to introduce a term in the Hamiltonian
that will set a preferred vertex valence and thus effectively
restrict the configuration space of the graph model from the space
of all possible graphs to the space of $d$-regular graphs.

The general form of this valence Hamiltonian should depend only on
the number of on edges attached to a given vertex, \be H_V = g_V
\sum_a \, f_a \left(\sum_b N_{ab}, v_0\right). \ee Here $g_V$ is a
positive coupling constant and $v_0$ is a free real parameter. The
function $f_a$ should be chosen such that its minimum occurs when
vertex $a$ has exactly $v_0$ on-links attached to it. The outer
sum over vertices $a$ indicates that all vertices in the graph
should have the same valence $v_0$ to minimize the total energy.

A specific choice of $H_V$ is \be H_V = g_V \sum_a \, e^{ p\,
\left(v_0 -\sum_b N_{ab} \right)^2} \ee where $p$ is another real
constant. The exponential is defined by its series expansion in
$p$: for example, for the $a=1$ term, \be \label{valpexp}
\begin{split} e^{ p\, \left(v_0\!-\!\sum_b \!N_{1b} \right)^2} \!\!\!= &1\! +\! p\left(v_0
-\!N_{12}-\!N_{13}-\!\cdots\! \right)^2 \\ &+
\frac{p^2}{2}\left(v_0 - \!N_{12} - \!N_{13} - \!\cdots
\!\right)^4 + O(p^3) \end{split} \ee The ellipses within the
parentheses stand instead of the summation over the other
$N_{1b}$.

Qualitatively, the effect of the valence Hamiltonian $H_V$ is to
set the preferred valence for a graph to $v_0$ and assign an
energy penalty for each vertex whose valence is $v\neq v_0$. It
will be important later that this penalty, for each vertex, scales
roughly with the exponential of the valence difference squared,
\be \label{deltaEVscale} \delta E_V \sim e^{p\,
\left(v-v_0\right)^2}.\ee The details of this scaling are
unfortunately somewhat complicated due to the normal-ordering
convention. Since the energy of a state is calculated using the
normal-ordering convention (\ref{EGnormal}) and this convention
implies relations such as (\ref{Numbersquared}), one has to be
careful when considering contributions from terms in which number
operators are raised to various powers: contributions such as
$N_{12} N_{12}$ in the expansion (\ref{valpexp}) give zero
regardless of whether the edge $N_{12}$ is on or off, and other
terms also disappear in this way. Despite these issues, it is
possible to check explicitly that once the energy is computed up
to sufficient order in $p$, the minimum of (\ref{valpexp}) is
indeed at $v_0$ and that the exponential scaling relation
(\ref{deltaEVscale}) holds. The remaining parts of this paper only
rely on the qualitative behavior of the valence term.

\subsubsection{Closed Paths}

The next terms that we consider involve powers of the matrix
$N_{ab}$ and depend on the number of closed paths present in a
graph.

At this stage we do not wish to introduce a bias for any
particular power, say $L=3$ or $4$ or $6$, corresponding to a
cycle length. Therefore, we would like to write a term for every
$L$ in the range $1 < L < \infty$. However, since we do want to
keep the dynamics of the model quasilocal, we would like to be
able to arrange, by adjusting some parameters or couplings, for
very high powers of $N_{ab}$ to be relatively unimportant.

There is more than one way to achieve this but we chose a
particular form which can be written down compactly as \be
\label{HB} H_B = \sum_a H_{B_a} \ee where $H_{B_a}$ is rooted at a
vertex $a$ and is given by \be \label{HBa} H_{B_a} \!\! = -g_B
\sum_b \; \delta_{ab} \, e^{r N_{ab}}. \ee Here $g_B$ is a
positive coupling and $r$ is a real parameter. The exponential is
defined in terms of a series expansion, \be e^{r N_{ab}} \, = \,
\sum_{L=0}^\infty \, \frac{r^L}{L!} \, N_{ab}^{(L)}. \ee Recall
from (\ref{Npowers}) that the operators of $N_{ab}^{(L)}$ return
the number of paths of length $L$ in the `on' graph that connect
vertices $a$ and $b$. When these powers of the number operator are
normal-ordered, paths that overlap become unimportant and only
nonoverlapping paths contribute. The sum over $b$ and the delta
function $\delta_{ab}$ in (\ref{HBa}) together ensure that only
closed paths are counted.

By (\ref{EGnormal}) then, this Hamiltonian assigns to the graph an
energy \be \label{EGpaths} E_B(G) \!= -
\!\sum_{a}\sum_{L=0}^\infty \, g_{B}(L) P(a,L) \ee that depends on
the number of closed paths $P(a,L)$ at each vertex $a$ of length
$L$. The ``effective'' coupling for each cycle length is given by
\be \label{gBeff} g_{B}(L) = \frac{r^L}{L!} \, g_{B}. \ee In
practice, terms with $L=0$ are uninteresting constants, terms with
$L=1$ vanish because there are no closed paths of that length, and
all terms with $L=2$ vanish because those closed paths are
necessarily overlapping. Hence the Hamiltonian $H_{B}$ starts
contributing at $L=3$.

A number of comments about this Hamiltonian are in order. First,
(\ref{HBa}) and (\ref{EGpaths}) both have an overall negative
sign. This indicates that a graph has lower energy the more cycles
it has. Since the system is finite, the Hamiltonian is bounded
from below and this negative sign does not create any problems.
Also, the energy associated per node is finite and constant for a
regular graph even when the number of vertices goes to infinity;
this is related to the second comment below.

Second, it is important to understand which of the various terms
contributing to $E_B(G)$ are most important. In graphs with a
large number of vertices, the number of long cycles at a vertex
$a$ is often larger than the number of short cycles, $P(a,L\gg1) >
P(a,L\sim 1)$. For certain classes of graphs, it is possible to
estimate the growth of $P(a,L)$ with $L$: for example, all
$v$-regular graphs have $P(a,L)$ bounded from above by a
polynomial of order $v^{L-1}$. However, the effective coupling
$g_B(L)$ multiplying $P(a,L)$ in (\ref{EGpaths}) falls faster than
any power for large $L$. Hence it is guaranteed that extremely
long cycles do not contribute significantly to the energy of a
graph. In this sense $H_{B_a}$ is a quasilocal operator that
contributes only a finite amount to the energy of a vertex.

For intermediate values of $L$, the situation is more subtle. The
effective coupling is maximized at a length $L^*$ for which \be
g_B(L^*) \,>\, g_B(L), \qquad \forall \, L\neq L^*. \ee This scale
depends solely on the parameter $r$. Another characteristic length
is $L_a^{**}$ defined for each vertex by \be g_B(L_a^{**}) \,
P(a,L_a^{**}) \,>\, g_B(L) \, P(a,L), \qquad \forall \, L \neq
L_a^{**}. \ee This second length is graph dependent. The lengths
that are relevant for determining the total energy assigned to a
vertex or a graph range from zero to some multiple of $L^{**}$.

Third and last, note that the parameter $r$ is raised to various
powers in (\ref{gBeff}). If this parameter is positive, so is the
effective coupling $g_B(L)$. If this parameter is negative,
however, the effective coupling $g_B(L)$ has a different sign for
cycles of even and odd lengths. In the latter case, even cycles
can lower the energy of a graph while odd cycles incur it a
positive energy penalty.

\subsubsection{Interaction Terms}

The previous terms $H_V$ and $H_B$ can be thought of as being
terms in a ``free'' graph model - they are eigenoperators of graph
states and do not change the linking structure between vertices. A
general graph model Hamiltonian might also have some ``interaction
terms'' which change the graph diagram of a graph state. As a
matter of principle, in fact, interaction terms are necessary in a
graph model because they define how a graph state can evolve from
one configuration to another.

\begin{figure}[t]
  \begin{center}
    \mbox{
      \subfigure[]{\includegraphics[scale=0.75]{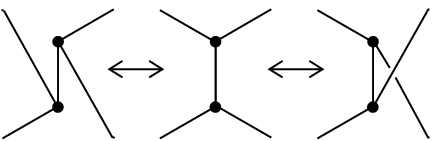}}
      \quad\;\;
      \subfigure[]{\includegraphics[scale=0.75]{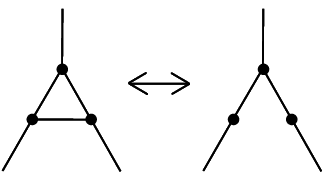}}
      }
    \caption{Interaction moves on graphs. (a) Exchange moves preserve the valence of each vertex. For convenience the move
    between the center and the left is called type I and the move between the center and the right is called type II. (b) Other
    moves can add or subtract edges, changing the valence of some nodes. This move is called type III.}
    \label{fig_moves}
  \end{center}
\end{figure}

One might think of many possible interactions for graphs. However,
we would like to impose a restriction of locality on the
interactions so that they affect only small local neighborhoods of
vertices. Some examples of such possible interactions are shown in
Fig. \ref{fig_moves}.

In terms of operators, the exchange interaction of Fig.
\ref{fig_moves}(a) can be formulated as \be H_{(exch)} =
g_{(exch)} {\sum_{abcd}}^{\,\prime} \; N_{ab} \left( a^\da_{ad} \,
a^\da_{bc} \, a_{bd} \,a_{ac} \right). \ee The prime on the
summation indicates that the vertices $abcd$ are assumed to be all
different. The presence of the number operator $N_{ab}$ in the
interaction term ensures that the exchange operation between
vertices $abcd$ takes place if and only if the link between
vertices $a$ and $b$ is on. Moves of type I and type II
distinguished in Fig. \ref{fig_moves}(a) are subcases of this
formula.

The addition and subtraction move of Fig. \ref{fig_moves}(b) can
be written as \be H_{(add)} = g_{(add)} {\sum_{abc}}^{\,\prime}
N_{ab} N_{ac} \left( a_{bc} + a^\da_{bc} \right). \ee Again, the
sum over $abc$ is assumed to be such that the considered vertices
are all different. The creation and annihilation operators in the
parentheses add or subtract an edge at $bc$ if and only if there
are already edges connecting $ab$ and $ac$.

It is possible to generalize these terms such that they exchange or
add links between vertices that are more separated from each other.

The couplings $g_{exch}$ and $g_{add}$ determine how likely the
interactions are to happen. In the next section, we mainly study
static or equilibrium configurations of links and therefore ignore
the interactions. The exchange moves will only play a role in the
discussion of perturbations.


\section{Large Graph Example}\label{s_largegraph}

Consider a system in which the number of vertices is very large,
for example $N\sim 10^{100}$ or $N\sim10^{1000}.$ The number $N$,
however large, is always thought to be finite. Technicalities and
physical interpretation of the limit $N\ra\infty$ are not
considered.

We begin by noting that regular lattices can be thought of as
special regular graphs in which some of the cycles correspond to
plaquettes. For example, the two-dimensional honeycomb lattice has
the same number of cycles supported by each pair of edges at each
vertex. In this section we thus study a graph model as defined in
Sec. \ref{s_graph_prelim} and ask whether a lattice with hexagonal
plaquettes can correspond to the graph state that minimizes the
energy assignment $E(G)$.

Since the hexagonal lattice is $3$-regular, a reasonable guess for
a Hamiltonian that might produce it is \be H \label{Htry1} = H_V +
H_{B} \ee with the preferred valence set to \be v_0=3 \ee and the
couplings set to \be \label{reg3} g_V \gg g_B, \qquad g_B = 1. \ee
The coupling $g_V$ has to be large to enforce the $3$-regularity
condition, while the normalization of $g_B$ is arbitrary. Since
the honeycomb has plaquettes of length 6, we consider values of
$r$ so that $L^*$ and $L^{**}$ are close to 6.

\subsection{Finding the ground state}

The ground state in this section is defined as the graph $G_0$ for
which $E(G_0)$ is smaller than $E(G)$ for any other $G$. A
discussion of when $G_0$ can be expected to be the optimal
configuration also from the point of view of statistical mechanics
is postponed until Sec. \ref{s_statmech}.

\begin{figure}[t]
  \begin{center}
    \mbox{
      \subfigure[]{\includegraphics[scale=0.7,viewport=0 -20 132
      120]{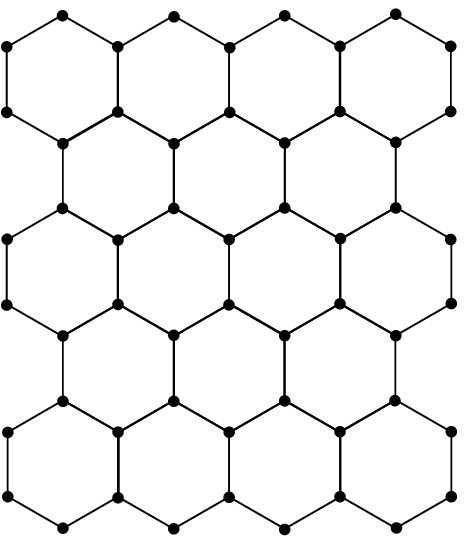}}\qquad
      \subfigure[]{\includegraphics[scale=0.7,viewport=0 -22 24
      120]{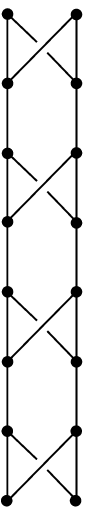}}\qquad
      \subfigure[]{\includegraphics[scale=0.7,angle=90]{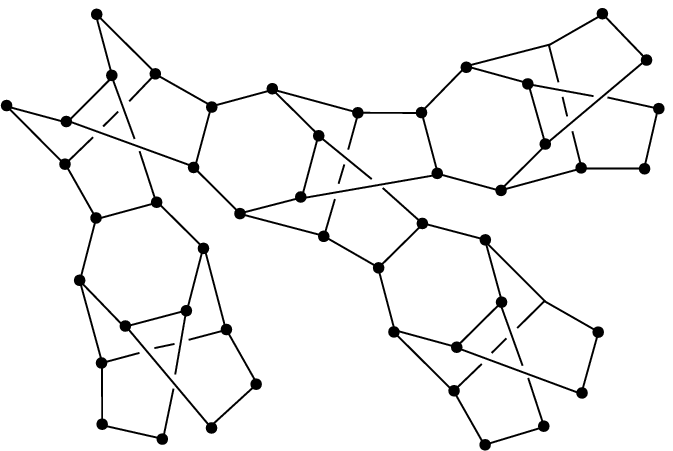}}
      }
    \caption{Sample $3$-regular graphs: (a) hexagonal lattice, (b) braided line, and (c) braided tree.}
    \label{f_val3}
  \end{center}
\end{figure}

Using the condition (\ref{reg3}) and the fact that the energy
penalty for a vertex to have valence different from $v_0$ grows
very rapidly, we focus attention on the class of $3$-regular
graphs and study primarily the effect of the cycle term $H_B$.
Since the honeycomb lattice has plaquettes of length 6, a first
attempt at choosing the parameter $r$ could be such to make
$L^{**}=6$, i.e. so that terms proportional to paths with $L=6$
contribute the most to $E(G)$. This can be quickly realized to be
unsuccessful as there are several $3$-regular graphs that have
more closed paths of length $6$ than the hexagonal arrangement;
some of these are shown in Fig. \ref{f_val3}.

In the graphs of Fig. \ref{f_val3}(b) and Fig. \ref{f_val3}(c),
many of the $6$-cycles have more than one edge in common. This
property causes these graphs to be effectively lower dimensional
than the hexagonal lattice: Fig. \ref{f_val3}(b), for example, can
be seen as one-dimensional on the large scale. A related
consequence of this property is that the numbers of long cycles in
graphs in Fig. \ref{f_val3}(b) and Fig. \ref{f_val3}(c) are lower
than in the case of Fig. \ref{f_val3}(a).

\begin{figure}[t]
  \begin{center}
        \includegraphics[scale=0.7]{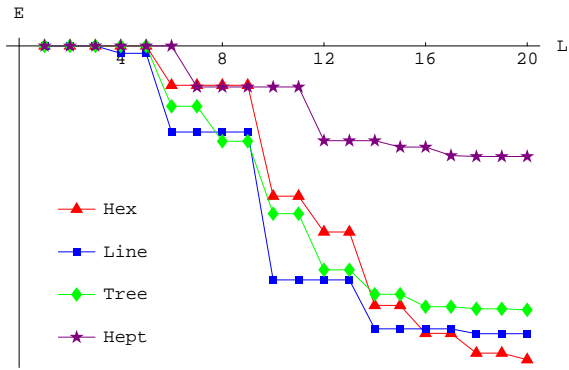}
    \caption{The value of the loop energy per vertex, in units of $g_B$,
    for some sample graphs including the flat hexagonal lattice as a function of cutoff length $L$. The parameter $r$ is set to $7.3$.}
    \label{f_hexwins}
  \end{center}
\end{figure}

It is impractical to count cycles of all possible lengths for each
of the candidate graphs. We know from the behavior of the loop
term, however, that this is not necessary as very long cycles do
not contribute much to the energy $E_B(G)$. Thus it is reasonable
to cut-off the sum over length in the definition of $E_B(G)$ at
some finite value of $L$. Figure \ref{f_hexwins} shows the energy
per vertex in each of the candidate graphs, plus a regular lattice
made up of heptagons, as a function of the cutoff length $L$.

The figure illustrates a number of interesting features. When the
cutoff length is taken to be small, the energy per vertex for the
braided graphs is lower than that of the hexagonal lattice. It is
only when cycles of lengths up to around 20 are considered that
the hexagonal lattice becomes the preferred configurations out of
the four candidates. The fact that all four lines tend to level
off at high cut-off lengths demonstrates that longer cycles become
increasingly less important.

Other interesting facts that can be seen from the figure relate
how much cycles of each length contribute to the whole loop energy
for each graph. For the case of the hexagonal lattice with $|r|
\sim 6,7$ such that $L^*\sim 6,7$, one finds that the cycles of
length $10$ contribute the most (the jump in height between $L<10$
and $L=10$ is greatest than the other ones) but also that cycles
of length $14$ are more important than cycles of length $6$, $12$,
or other lengths $L>14$. For the same values of $r$, the most
important cycles in the braided line [Fig. \ref{f_val3}(b)] have
lengths $10$, $6$, $14$ - the relative importance of cycles of
length $6$ and $14$ are switched compared to the hexagonal
lattice. In the overall picture, these nuances do not seem very
important but they do indicate that the dependence of $E(G)$ on
the graph structure is nontrivial.

Figure \ref{f_hexwins} shows that the hexagonal lattice is
preferred over the other graphs once cycles of all lengths are
considered, when the parameter $r$ is $r=+7.3$. In fact, the same
empirical conclusion, when comparing the four lines in the plot,
can be reached for other values of $r$, \be \label{rthreshold} |r|
\gtrsim 7.1. \ee For $r$ close to the lower bound
(\ref{rthreshold}), lengths up to $L=20$ allow one to compute
$E(G)$ up to $1\%$ for the hexagonal and better for the other
lattices, and the energy differences between the hexagonal and the
braided line and braided tree, are $3\%$ and $16\%$, respectively.
The energy difference between the hexagonal lattice and the
heptagonal lattice is much larger as seen in the figure.

The data used to plot the ``hex'' line in Fig. \ref{f_hexwins} is
obtained by counting cycles in the two-dimensional flat hexagonal
lattice. If an arrangement of hexagons as in Fig. \ref{f_val3} is
wrapped in a tube or torus, then the energy per node can be set up
to be lower than that shown in the plot even by a factor of 2 if
the circumference of the tube is about 8 edges. This large
discrepancy is due to cycles that wind around the tube and lower
the energy relative to the flat configuration. In any case, it
seems that among the various examples considered, it is a locally
hexagonal tube that corresponds to the lowest energy state of the
system. We stress again that the contribution coming from a wide
range of path lengths must be considered in order to arrive at
these observations.

A general proof of the statement that a locally hexagonal lattice
is the true ground state of the model is at this moment beyond
reach and so we can only phrase it as a conjecture. Because the
number of $3$-regular graphs with $N$ vertices is very large, a
brute force search for the ground state would be a very
computationally intensive task. In general, any approach to
finding the true ground state would be complicated by the
necessity to consider very long paths in the analysis. In the
above discussion, we focused attention on some candidates which
have a large number of cycles of lengths close to 6 as these could
have been considered as possible counterexamples to the
conjecture, and showed that they actually have higher energy.

Further evidence that the hexagonal lattice is at least a local
minimum in the model's energy landscape is presented in the next
section.

\subsection{Graphs Above the Ground State \label{s_perturbations}}

Based on the evidence above, one can try to proceed by assuming
that the hexagonal lattice is indeed the minimal energy
configuration for a given set of couplings. This lattice
configuration will be hereafter called the reference lattice. (For
simplicity we consider the flat hexagonal lattice, not the tube,
as the reference lattice.) It is interesting to consider graph
states that are close to this reference configuration and to
check, in a perturbation theory manner, that the reference lattice
is at least a local minimum in the state space. Given that the
local minimum property is confirmed, this procedure should also
provide information about the spectrum of low-energy graph
excitations.

\begin{figure}[t]
  \begin{center}
      \mbox{
      \subfigure[]{\includegraphics[scale=0.75]{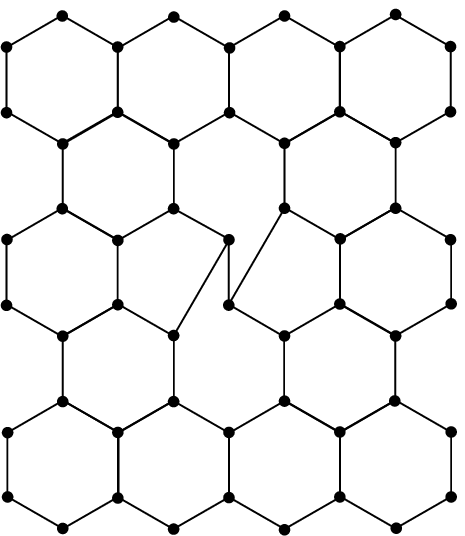}}
      \subfigure[]{\includegraphics[scale=0.48]{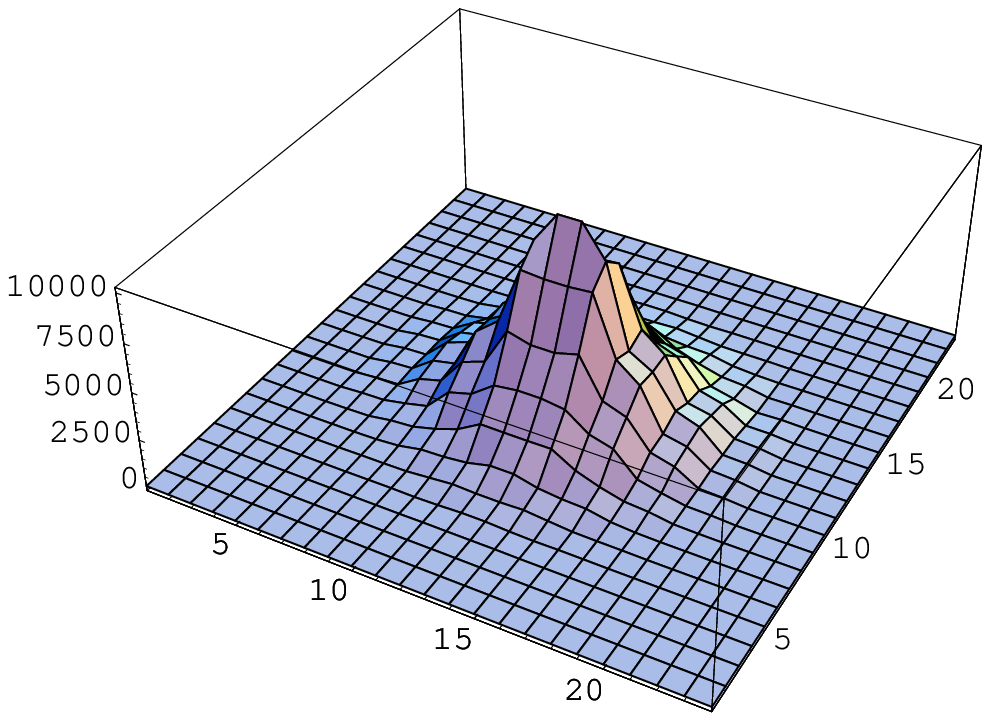}}
      }
    \caption{(a) The hexagonal lattice with a defect
of type I. (b) The plot shows the energy differences at points in
the hexagonal lattice relative to the reference lattice. The
vertical axis is in units of $g_B$. }
    \label{f_perturb1}
  \end{center}
\end{figure}

A possible type of perturbation around the reference state can be
done by applying one of the exchange moves shown in Fig.
\ref{fig_moves}. After one such move at an arbitrary location in
the reference graph, one obtains a new state as shown in Fig.
\ref{f_perturb1}(a). This state is still $3$-regular and therefore
its associated energy with respect to the valence term $H_V$ is
unchanged. The cycle structure changes since some of the closed
paths of length $6$ near the defect get replaced by closed paths
of lengths $5$ and $7$. The distribution of longer paths is also
affected. These structural changes alter the energy assignment to
many vertices, also ones that are not immediately close to the
defect. The total change in energy (relative to the reference
lattice) due to the defect can then be defined as the sum of the
energy changes of all the vertices.

With $r=7.1$, a parameter consistent with the earlier arguments
for the reference hexagonal lattice being the minimal energy
state, the total energy difference turns out slightly positive for
this deformation. This is despite some vertices actually
experiencing a local energy decrease. For $r=-7.1$, the total
energy difference is decisively positive as can be readily
understood by noting that the defect decreases the cycle count at
even lengths and increases the cycle count at odd lengths, both of
which correspond to inflicting an energy penalty. A plot of the
energy difference at each vertex, using $r=7.1$, is shown in Fig.
\ref{f_perturb1}(b). In the plots, half of the points shown on the
square grid correspond to the actual energy differences computed
at the vertices of the hexagonal lattice while half are evaluated
from the former by linear interpolation.

\begin{figure}[t]
  \begin{center}
    \mbox{
      \subfigure[]{\includegraphics[scale=0.75]{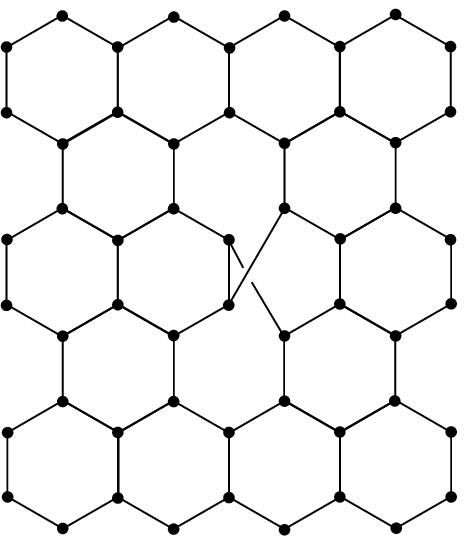}}
      \subfigure[]{\includegraphics[scale=0.48]{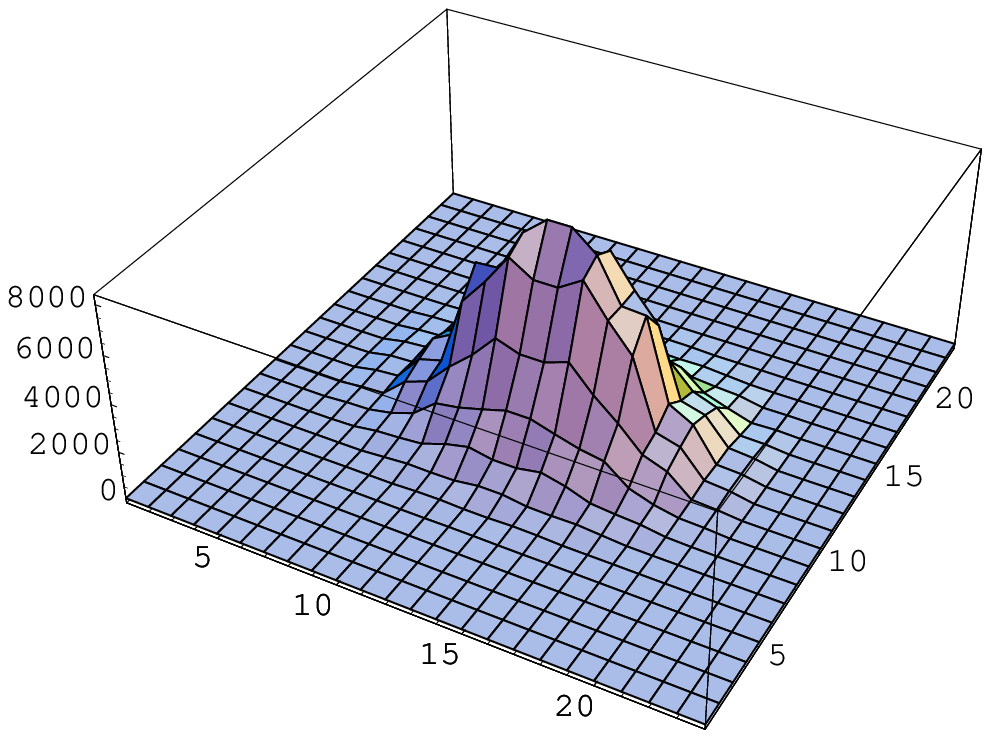}}
      }
    \caption{(a) The hexagonal lattice with a defect
of type II. (b) The corresponding energy difference plot. The
vertical axis is in units of $g_B$. }
    \label{f_perturb2}
  \end{center}
\end{figure}

A different perturbation can be obtained by applying an
interaction of type II from Fig. \ref{fig_moves} to the reference
graph. This gives a new configuration shown in Fig.
\ref{f_perturb2}(a). The corresponding energy difference plot is
shown in Fig. \ref{f_perturb2}(b). It is again computed using
$r=-7.1$. The shape of the plot in Fig. \ref{f_perturb2}(b) is
slightly different from the previous case, but still shows
unambiguously that the overall energy difference due to defect is
positive. In the case of this defect, the choice of negative $r$
is necessary because a positive value of $r$ actually decreases
the total energy.

From these perturbations, we learn that the reference lattice is
stable under deformations when the parameter $r$ is negative. With
negative $r$, therefore, the reference lattice is a local minimum
in the energy landscape and thereby a sensible candidate for the
ground state of the system. The fact that $r$ must be negative
implies that even and odd cycles in the model have an effective
coupling $g_B(L)$ of different sign and thus have quite distinct
physical effects.

The two deformed lattices considered have energies that are on the
order of $10^6g_B$ higher than the reference lattice and represent
two distinct excited states of geometry. They are probably not the
two lowest-lying states, but they are ones that come about by
disturbing the reference lattice with minimal local moves.
Perturbations from the type III move of Fig. \ref{fig_moves} can
be studied in a similar manner.

While a full description of the spectrum of excitations is still
missing, it is already possible to say a few things about the
spectrum. At energies much higher than those corresponding to the
two lattices with defects but still smaller than the coupling
$g_V$, excited states can be expected to be graphs in which all
vertices have $3$ edges but in which the cycle structure is much
different from the reference lattice. At energies beyond $g_V$,
excited states can also appear in which some vertices have higher
(or lower) than the preferred valence. The physical consequences
of such perturbations on Ising systems have already been studied
\cite{Wan:2005ip}. In the context of our model, characterizing
these states is again difficult because the new edges
corresponding to higher valence create many new cycles which
contribute through the loop term. In any case, at sufficiently
high energy the states of the system become highly irregular and
cannot be expected to be interpreted as lattices with defects.
These states characterize the disordered phase described in the
Introduction.

\subsection{Statistical Mechanics \label{s_statmech}}

In this section, we continue discussing the stability of the
reference lattice state. We consider the thermodynamic definition
of the free-energy $F=E-TS$ as the relation between the energy $E$
and the entropy $S$ at a given temperature $T$. To check that the
reference lattice is a stable configuration, we consider the
variation \be \delta F = \delta E - T \delta S \ee after small and
large local perturbations. In this context, we take locality to
refer to the emergent locality of the reference lattice.

To start, consider a local subgraph of the reference lattice
consisting of $1 \ll n \ll N$ vertices. Then, consider single
perturbations of the type shown in Fig. \ref{fig_moves}(a) and
discussed in Sec. \ref{s_perturbations}. The corresponding energy
shifts $\delta E$ were computed and displayed in Fig.
\ref{f_perturb1}(b) and Fig. \ref{f_perturb2}(b). For the present
purpose, it is sufficient to note that these single perturbations
contribute $\delta E \sim const$. Since the perturbation move can
be applied to any one of $O(n)$ links in the subgraph, the entropy
change associated with a perturbation is $\delta S \sim \ln n$.
Putting these elements together, \be \label{deltaF1} \delta F \sim
constant - T \ln n. \ee At finite temperature, $T\neq 0$ and large
$n$, the change in free-energy $\delta F$ can be negative. This
means that small defects are likely to be present in the system at
nonzero temperatures. This is not a negative result and may
possibly lead to specific signature of the model that can be
compared with experiment. The number density of such defects
depends on the scale of $\delta E$ and the temperature $T$.

Next, consider changing the graph by applying several exchange
moves to a region of the graph composed of $m$ vertices that are
close together, with $m<n$. The change in entropy $\delta S$
associated with this change is still proportional to $\ln n$. The
energy shift is more difficult to estimate as a function of $m$
because it very much depends on what the perturbation is. However,
since it is assumed that the perturbation affects $m$ vertices, it
is reasonable to take $\delta E$ to be at least of order $m$,
$\delta E \sim \Omega(m)$. Thus the change in free energy \be
\label{deltaF2} \delta F \sim \Omega(m) - T \ln n \ee is positive
for sufficiently large $m$. This indicates that the reference
lattice is stable against large local perturbations even at finite
temperature: the reference lattice describes a stable
thermodynamic phase.

Finally, consider a similar setup to the one just discussed, but
instead of applying several exchange moves, consider changing the
valence of a region of $m<n$ vertices in the reference lattice
from $3$ to $v$. Vertices with valence $v\neq 3$ will incur an
energy penalty due to the valence term $H_V$. However, since the
subgraph with higher valence contains a larger number of cycles
than the reference lattice, the loop term $H_B$ will decrease the
energy somewhat. For a $v-$ regular graph, the number of cycles of
each length is bounded by \be \label{Pbound} P(a,L) < c v^{L-1}
\ee where $c$ is some constant. This can be used to pose a bound
on the effect of the loop term as \be \sum_{L} g_B(L) \, P(a,L) <
g_B \sum_L \frac{r^L}{L!} c v^{L-1} \sim g_B \frac{e^{rv}}{v}. \ee
The overall energy difference, for each affected vertex, scales
thus as \be \label{deltaEa} \delta E_a = g_V \, e^{p\, (v-v_0)^2}
- g_B \, \frac{e^{rv}}{v}. \ee As long as this quantity is
positive, the higher valence droplet has higher energy per vertex
than the reference lattice. From here, using similar reasoning as
in the case of the other type of perturbation, we can conclude
that \be \label{deltaF3} \delta F \sim \Omega(m) - T\ln n \ee just
like in (\ref{deltaF2}). The reference lattice is thus also stable
against changes in the valence.

This analysis can be compared to similar heuristic arguments that
are used to show how dimensionality, range of interactions, and
type of interactions determine whether a system of spins on a
fixed lattice can exhibit order-disorder phase transitions
\cite{Sewell}. For the simple spin systems on a lattice, such
arguments can be made precise \cite{SimonSokal}. Whether a similar
level of rigor can be achieved for the graph model system is still
unknown.

\section{Extensions with More Degrees of Freedom \label{s_moredof}}

Whereas the model of Sec. \ref{s_model} had a minimal Hilbert
space on each of the edges, one may also be interested in models
that contain more degrees of freedom. In this section we describe
how this could be done and explain how such more complex models
can connect to quantum field theories, including quantum gravity.

Again, the goal in this section is to define more complex models
by altering the Hilbert space $\HH_{edge}$ associated to each edge
in the complete graph. We still require that the new $\HH_{edge}$
contain a state $|0\rangle$ that can be interpreted as the
physical link between two vertices being off. But now, instead of
creating an on state by acting with a creation operator $a^\da$,
we introduce a set of such operators ${a_s}^{\!\!\!\da}$ labeled
by an index $s$ chosen from a set of integers. We also introduce
corresponding annihilation operators $a_s$. As before, these
operators are set to obey fermionic anticommutation relations \be
\label{newanticommute} \{a_s, {a^\da_{s^\prime}} \} = \delta_{s
s^\prime}. \ee All other anticommutators at each edge vanish.

For concreteness, we here focus on $s=\{1,2,3\}.$ The Hilbert
space of the new edge is the span of all possible states that can
be constructed by acting with the $a_s^{\da}$. It is \be
\label{newHHedge} \HH_{edge} = \mathrm{span}\{ |\,0\rangle, \,
|\,1_1\rangle, \, |\,1_2\rangle, \, |\,1_3\rangle \}, \ee and the
states \be |\,1_s\rangle = a_s^{\da} |\,0\rangle \ee are all
interpreted as on states. It follows from the anticommutating
nature of the $a^\da_s$ that states with multiple particles cannot
exist.

The difference between the edge Hilbert space (\ref{newHHedge})
and the old one (\ref{HHedge}) is that there are now multiple on
states that can be distinguished by an internal label $s$. The
total Hilbert space for this extended model is defined as in
(\ref{HHtot}) and can still be decomposed according to
(\ref{HHtotsum}). However, the spaces $\HH_G$ in the tensor sum
decomposition are here no longer zero-dimensional but reflect the
internal degrees of freedom of the on links. Thus, the spaces
$\HH_G$ have now room for interesting physics. In what follows,
operators that rotate between these internal states are used in a
Hamiltonian to describe matter degrees of freedom propagating on a
dynamically selected background graph.

In order to connect with the original quantum graphity model
\cite{graphity1}, consider relabeling the states of
(\ref{newHHedge}) so that \be
\begin{split} \HH_{edge} &= \mathrm{span} \, \{ |\,0, 0 \rangle,\; |\,1, -1\rangle, \;|\,1,
0\rangle, \; |\,1,\,+1\rangle \} \\ &= \mathrm{span} \, \{ |\,j,\,
m\rangle \} \end{split} \ee so that the off state has $j=0$ and
$m=0$, and the on states have $j=1$ and $m=0,\pm 1$. There is a
clear analogy between this space and the Hilbert space of a spin-1
particle and hence it is natural to introduce an operator $M$
which has the states $|j, \, m\rangle$ as eigenstates, \be M\,
|\,j, \, m\rangle = m\,|\,j,\, m\rangle, \ee and operators $M^\pm$
that change the internal $m$ labels, \be
\begin{split}
M^+ \, |j, \, m\rangle &= \sqrt{(j-m)(j+m+1)} \, |\,j, \, m+1\rangle \\
M^- \, |j, \, m\rangle &= \sqrt{(j+m)(j-m+1)} \, |\,j, \,
m-1\rangle.
\end{split} \ee
These are the familiar operators of angular momentum (although the
$M$ and $M^\pm$ operators are sometimes called $J_z$ and $J^\pm$
instead). These three operators form a closed algebra among
themselves \be [M^+, \,M^-] = M, \qquad [M,\,M^{\pm}] = \pm
M^{\pm} \ee and annihilate the $|\,0,\, 0\rangle$ state, \be
M\,|\,0,\, 0\rangle = M^\pm \, |\,0,\, 0\rangle = 0. \ee The
formulation of $M$ and $M^\pm$ in terms of the creation and
annihilation operators $a^\da_s$ and $a_s$ is not needed in what
follows. Similarly as operators $N_{ab}$ of the original model,
the operators $M_{ab}$ and $M^\pm_{ab}$ that act on each edge can
also be organized and understood as being attached to an adjacency
matrix. Their powers also contain information about the closed and
open paths of a graph state.

A Hamiltonian for a model with this edge structure can be written,
for example, as (graph interaction terms are not shown) \be H =
H_V + H_B + H_C +H_D + H_\pm \ee where $H_V$ and $H_B$ are the
same as in Sec. \ref{s_model} and the other terms are \be
\label{HC} H_{C} = g_C \sum_a\left(\sum_b M_{ab}\right)^2, \ee \be
\label{HD} H_D = g_D\sum_{ab}M^2_{ab},\ee \be \label{Hpm} H_{\pm}
=-\sum_{\rm cycles}g_{\pm} \left(L\right)\prod_{i=1}^L M^{\pm}_i.
\ee Here $g_C, g_D$, and $g_\pm$ are additional positive
couplings. In the $H_\pm$ term, referred to as the loop term
below, the product is taken around a cycle of length $L$ (i.e.,
consisting of $L$ edges) and with alternating raising and lowering
operators: \be \label{loopoperator} \prod_{i=1}^L
M_i^\pm=M^+_{ab}M^-_{bc}...M^+_{yz}M^-_{za}. \ee Since this
product contains an equal number of raising and lowering
operators, the loop operator is naturally restricted to act on
cycles of even length. The coupling $g_\pm(L)$ is \be g_\pm(L) =
\frac{r^L}{L!} g_\pm. \label{eq:BL} \ee Note the similarity of
this coupling function to that of (\ref{gBeff}) in the loop term
$H_B$. Actually, the original quantum graphity was written only
with the $H_\pm$ term, without the $H_B$ term. Both are included
here because this makes it easier to separate the graph-forming
role of $H_B$ from the matter propagation role of $H_\pm$.

By the arguments of Sec. \ref{s_model}, we assume that the ground
state of the system at very low temperatures is a $3-$regular
graph with hexagonal symmetry. Since the new terms of the
Hamiltonian contain only $M$ and $M^\pm$ operators and not $a_s$
and $a_s^\da$ operators by themselves, they do not change the
linking configuration. At low temperatures, therefore, we can
consider the base graph to be frozen in a hexagonal configuration
and discuss the action of $H_{CD}$ and $H_\pm$ on this background.
Then, the terms of (\ref{HC}), (\ref{HD}) and (\ref{Hpm}) reduce
to a model of string nets \cite{LevinWen}. We briefly describe the
expected physics.

Since the loop Hamiltonian $H_{\pm}$ does not commute with $H_{C}$
or $H_{D}$, the eigenstates of the full Hamiltonian will generally
be superpositions of states involving different $m$
configurations. Nonetheless, an intuition for the model can be
developed by first describing the eigenstates of the $H_C+H_D$
terms alone, and then considering the effect of the loop term. The
ground state of $H_C+H_D$ consists of all links having $m=0$. When
$g_C \gg g_D$, low-energy excited states appear as closed chains
of links on which the $m$ variables have alternating values $m=+1$
and $m=-1$. These excitations are called strings and their energy
above the ground state is proportional to the coupling $g_D$ times
their length (number of edges.) Thus $g_D$ can be thought of as a
string tension. The coupling $g_C$ can instead be related to the
mass of pointlike particles \cite{LevinWen}.

Given a graph with all on edges labeled by $m=0$, a loop operator
(\ref{loopoperator}) acts as to create a closed string of
alternating $m=+1$ and $m=-1$ edges (a loop operator cannot create
open strings.) These closed strings acquire tension through the
$g_D$ term. However, since the sign of the $g_\pm$ term is
negative, the overall energy of the state may either increase or
decrease as a result of string creation and so there is the
possibility of two distinct scenarios. In one scenario, the
tension in a string is greater than the contribution from the loop
term, so the overall effect of creating a string is to increase
the energy of the system. If this is the case, then the string
represents an excited state over the vacuum in which all $m$
values are set to zero. The second scenario is the one that we
will be mostly interested in. If the tension is small compared to
the contribution from $H_{\pm}$ so that creating a string
decreases the energy, then the creation of the string actually
lowers the energy and indicates that the original configuration
cannot be the ground state. Instead, the true ground state
consists of a superposition of a large number of strings - a
string condensate. We should note that because the graph has a
finite number of vertices and the $m$ values on each edge only
take three possible values, the Hamiltonian is bounded from below.
The characterization of the string-condensed ground state is
difficult but its excitations are expected to be that of a $U(1)$
gauge theory \cite{LevinWen} since the Hamiltonian is close in
form to the Kogut-Susskind formulation of lattice gauge theory
\cite{Kogut}. The two main differences between this model and the
original string-net condensation model proposed by Levin and Wen
\cite{LevinWen} are that in the present case the background
lattice is dynamical and has hexagonal rather than square
plaquettes.

Another possibility for incorporating matter and indeed
gravitational degrees of freedom in the graph-based model that is
worth mentioning is via the approach of algebraic quantum gravity
\cite{Giesel:2006uj}.

\section{Comparison to Other Graph Dynamics}\label{s_graphtheory}

In this section we describe the quantum graphity model of Sec.
\ref{s_model} in graph-theoretic terms and compare its dynamics
under the $H_V+H_B$ Hamiltonian to some common graph processes
discussed in the literature. In particular we are interested in
modeling the high-temperature to low-temperature transition with a
mechanism acting on the graph associated to the system. A
\emph{graph process} can be defined by taking into account two
ingredients: an \emph{initial graph} and a set of \emph{graph
operations}. The process consists of applying in sequence the
graph operations from the set. In this way, the initial graph is
gradually transformed into other graphs, according to the
operations used.

In our scenario, the initial graph can be any graph with a very high
density of edges, because this is what we expect to be the likeliest
state of the system when the temperature is very high. For
simplicity, however, we can take this initial state to be $K_N$.
It is intuitive, and somehow simplest, to consider a unique
operation, which, in our case, is the deletion of edges.

We denote by $G=(V,E)$ a graph with set of vertices $V(G)$ and set
of edges $E(G)$. A graph $G=(V,E)$ is \emph{d}-\emph{regular} if
$d(i)=d$ for all $i\in V(G)$, where $d(i):=|\{j:\{i,j\}\in
E(G)\}|$ is the \emph{degree} of the vertex $i$. In what follows
we will mainly focus on regular graphs with small degree, and more
specifically, $3$-regular (also called \emph{cubic}) graphs. By
$H\subset G$, we mean that $H$ is a graph with $V(H)\equiv V(G)$
and $E(H)\subset E(G)$. One may interpret $H$ as obtained from $G$
by deleting edges of $G$ but keeping all of its vertices. We
consider a family of graphs $\{G_{i}\}$, such that $G_{0}=K_{N}$
and $G_{k}\subset \cdots \subset G_{1}\subset G_{0}=K_{N}$. The
dynamics induced by the Hamiltonian $H_{VB}$ requires the graph
$G_{k}$ to satisfy the following two conditions: that $G_{k}$ is
$v_{0}$-regular and that $E_{B}(G_{k})<E_{B}(G_{i})$, for all
$i=0,...,k-1$, according to Eq. (\ref{EGpaths}). Note that these
conditions are \emph{local} at the level of the vertices, that is
both conditions can be verified by looking at the single vertices
of the graph. The fact that $G_{k}$ needs to be $v_{0}$-regular
can be easily verified and enforced. The fact that $ E_{B}(G_{k})$
is small depends on the cycle structure of $G_{k}$. Devising a
graph process to control the number of cycles having different
lengths for each vertex does not appear to be an easy task.

We consider the dynamics towards the ground state of $H_V+H_B$ as a
process that transforms $G_{i}$ into $G_{i+1}$ by deleting edges of
$G_{i}$. It is evident that $E_{B}(G_{i})$ is small, when $G_{i}$
has a relatively large number of cycles of length between zero and
some $L_{\max }$ \emph{not much larger} than $L^{\ast }$.

The first natural idea is to consider random graphs (see,
\emph{e.g.}, \cite{ka}). The best known of such models are the
Erd\"{o}s-Renyi random graph and the uniform random graph. The
\emph{Erd\"{o}s-Renyi random graph} $G(N;p)$ on a set of $N$
vertices is obtained by drawing an edge between each pair of
vertices, randomly and independently, with probability $p$. The
featured randomness in assigning edges does not insure to obtain a
$v_{0}$-regular graph. Therefore, these models do not satisfy our
first requirement.

In a \emph{random graph process}, one starts with $N$ vertices and
inserts edges one at a time at random. While this process does not
guarantee to generate a $d$-regular graph either, this model is
more pertinent to our setting since adding edges starting from $N$
vertices and no edges at all is conceptually equivalent to
deleting edges starting from the complete graph $K_{N}$. However,
when $d$ is relatively small the behavior of the latter model is
not easy to analyze given that it needs to run for a large number
of steps. These are not the last words, since there are
well-defined models of random regular graphs \cite{Wormald}.
Indeed, the so-called $d$\emph{-process} is similar to the random
graph process, but the degrees of the vertices are bounded above
by a constant $d$. This process gives a $d$-regular graph with
probability tending to $1$ as the number of vertices tends to
infinity. Note that a $d$-process does not consider at all the
number of cycles for each vertex, our important second
requirement. What can we say about this point? Let us focus on the
case $v_{0}=3$. It is known that the probability $Pr(t)$ that a
graph chosen at random from the set of all cubic graphs on $N$
vertices contains exactly $t$ cycles of length $L$ (where $t$ is
fixed) goes to \be Pr(t) \,\sim \,\frac{e^{-2^{L}/2L}}{t!}\ee as
$N\rightarrow \infty $. Also, the expected number of cycles of
length $L$ in a random cubic graph on $N$ vertices goes to
$2^{L}/2L$ as $N\rightarrow \infty $. In our model we must take
into account the term $H_B$ in the Hamiltonian that depends on the
cycles. Because of this term, the graph associated to the ground
state of the Hamiltonian needs to have a relatively large number
of short cycles. The above observation about the cycle structure
in $d$-processes does not reflect this behavior. It follows that
$d$-processes do not seem to be good candidates to implement the
dynamics suggested by our model. Another reason supporting this
statement comes from the diameter. For a $d$-regular graph $G$ on
$N$ vertices, \be diam(G)\leq 1+\left\lceil \log
_{d-1}((2+\epsilon )dN\log N)\right\rceil, \ee with probability
tending to $1$ as $N\rightarrow \infty $. Since $N$ is very large
in our context, we can consider the above formula as a good
approximation. Note that the behavior of diam$(G)$ exhibits an
interesting cutoff phenomenon: $diam(G)$ increases very slowly
when $d\lesssim 10$ but rises quickly for $d\gtrsim 10$.
Conjecturing that the graph associated to the ground state of our
model has a small Hausdorff dimension $\delta $, the diameter
should be proportional to $N^{1/\delta }$, and this is much larger
than the diameter of a $d$-regular random graph.

In addition to the above reasoning, we can still observe that
random regular graphs can play a role in our mode. By taking
$g_{B}=0$, from a simple statistical mechanics argument suggests
that the probability of a vertex having degree $v$ is \be Pr
(v\neq v_{0}) \, \propto \, \exp (-\beta e^{(v-v_{0})^{2}}).\ee
When $N$ is very large, this can be considered as a good
approximation, despite the fact that the probability cannot be
taken independently for each vertex, given that increasing the
degree to one vertex implies increasing the degree of another one.
Also, when $g_B=0$, the cycles structure does not play a role in
determining the ground state. So, in this extremal case with
$g_{B}=0$, we can expect that the graphity model gives rise to a
$d$-regular random graph.

Finally, it is worth commenting on scale-free graphs which have
been widely discussed in the literature, also in the context of
quantum gravity \cite{Wan:2005ip,Requardt:2003qy}. A
\emph{scale-free graph} is a graph in which the degrees $d(v)$ of
vertices $v$ exhibit the Yule-Simon distribution  \be Pr
[d(v)=k]\sim k^{-\gamma }.\ee The exponent $\gamma$ is often in
the range $\gamma \in \lbrack 1,3]$. This means that a scale-free
graph has a few vertices with very high degree and many vertices
with very small degree. Because of the valence term $H_V$ in our
Hamiltonian which gives a high-energy penalty $e^{(v-v_{0})^{2}}$
to vertices with different degree from $v_0$, it is implausible to
have vertices with very large degree at low temperature in our
model.

It thus appears that many known ways to generate graphs cannot
reproduce the features implemented through the Hamiltonian of our
model. A graph process that would successfully reproduce the
dynamics of the Hamiltonian would necessarily have to involve a
cost function that would preferentially create $d-$regular graphs
with a large number of cycles of prescribed lengths. Defining such
a cost function in a plausible way is difficult. The cost function
in principle should take into account a value associated to each
edge. A candidate process could be one that carries on by greedily
deleting edges in agreement with the function. Each edge has a
cost and at each time-step the edge with the smallest cost is
deleted. The cost of each edge depends on the number of cycles of
prescribed lengths that will be in the graph after the deletion of
the edge. The cost function needs to be updated at each time-step,
since the deletion of a single edge implies a possibly large
variation on the number of cycles in which other edges are
contained. This observation suggests that implementing the
behavior induced by the Hamiltonian, with the use of a cost
function of pure combinatorial nature, is highly expensive from
the computational point of view, and it is possibly ill-defined.
For this reason, a mathematical description of a graph process
mimicking our dynamics is elusive.

\section{Discussion \label{s_discussion}}

Quantum graphity is an explicit model for geometrogenesis, with
locality, translation symmetry, \emph{etc.}, being properties of
the ground state. In Table \ref{t_summary}, we have summarized the
properties of quantum graphity at high and low temperatures.

At high temperature, the graph representing the state of the
system is highly connected and has diameter close to 1. There is
no notion of locality, as most of the Universe is one-edge
adjacent to any vertex. Said differently, there is no notion of a
subsystem, in the sense of a local neighborhood, since the
neighborhood of any vertex is the entire $K_N$. The microscopic
degrees of freedom are the $j$ and $m$ labels.

At low temperature, the graph has far fewer edges than $K_N$.
Permutation invariance of the state breaks to translation
invariance. Subsystems can be defined as subgraphs of the ground
state or, better, as the emergent matter excitations, and the
dynamics of the emergent matter is local by correspondence with
lattice field theory \cite{LevinWen,Kogut}. Once subsystems are
present, {\em internal geometry} can be defined. This is the
relational geometry that is the only physically meaningful notion
of geometry, and hence time, for observers inside a system. The
significance of internal time and its relation to general
relativity has been discussed extensively, for example, in
\cite{LQG,OD}.

\begin{table}[htdp]
\begin{tabular}{ll}
$\;\;\;$High-$T$&$\;\;\;$Low-$T$\\
\includegraphics[scale=0.8,viewport=-10 -26 120 120]{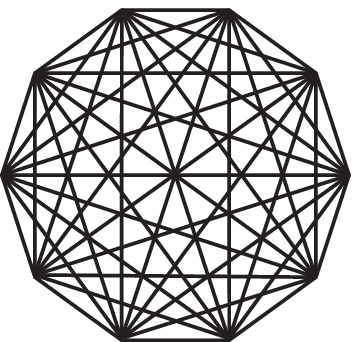}
&
\includegraphics[scale=0.75,viewport=-8 0 120 120]{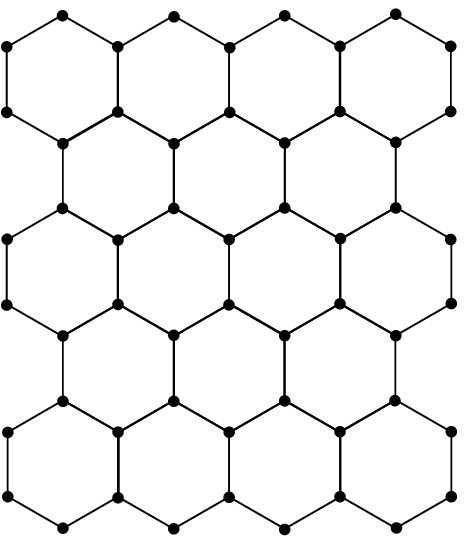}
\\
$\bullet$ Permutation symmetry&
$\bullet$ Translation symmetry\\
$\bullet$ No locality&$\bullet$ Local\\
$\bullet$ Relational&$\bullet$ Relational\\
$\bullet$ Diameter $\simeq 1$& $\bullet$ Large diameter\\
$\bullet$ $\sim N-$dimensional&$\bullet$ Low-dimensional\\
$\bullet$ No subsystems&$\bullet$ Lubsystems\\
$\bullet$ External time& $\bullet$ External and internal time\\
$\bullet$ $(j,m)$&$\bullet$ Matter + dynamical geometry\\
\end{tabular}
\caption{\label{t_summary} The two phases of quantum graphity. The
couplings are $g_V\gg g_B, g_\pm \gg g_C,$ and $g_D.$ The
parameters are $v_0, p,$ and  $r$.}
\end{table}

On a more technical level, the model is written in terms of
fermionic operators acting on a large Hilbert space based on a
complete graph. A term in the model Hamiltonian ($H_B$) that
detects the cycle structure of a graph is crucially defined using
powers of a quantum adjacency matrix and the normal-ordering
convention.

While our definition of $H_B$ has the properties that we would
like to implement in a graphity model, we find it interesting to
point out, as a side remark, that other choices could also be
made. For example, there is some interest in the graph theory
literature in a relation counting loops (in our notation) of the
form \be E_\zeta(G) = \sum_{L=1}^\infty \, \frac{r^L}{L} \,
\widetilde{P}(L), \ee where, in contrast to (\ref{EGpaths}) of our
model, the denominator on the right-hand side is $L$ rather $L!$
and the object $\widetilde{P}(L)$ denotes the number of prime
geodesics in a graph $G$ rather than the number of nonoverlapping
closed paths. This quantity $E_\zeta(G)$ is related to the Ihara
zeta function of a graph; we refer to the literature for more
details \cite{Terras} and finish this side remark by saying that
we have not looked at the behavior of $E_\zeta(G)$ in a graphity
model.

Besides the definition of the Hamiltonian term $H_B$ that depends
on quantum mechanical features, our study of the model is limited
to stationary states and is thus mostly classical. We find
evidence to support our conjecture that, for the given set of
parameters, the hexagonal lattice, possibly wrapped into a tube or
a torus, is a good candidate for the ground state of the model: it
has lower energy than other regular graphs with a large number of
cycles of length 6, and it corresponds to at least a local
free-energy minimum as found by heuristic arguments looking at
small and large local perturbations. The last argument relies on
the notion of emergent locality and the restriction of the
possible interactions in the graph to moves that act on small
subgraphs. These arguments could be extended to other ground state
lattices in different regions of the parameter space.

There are a number of important open questions. It would be useful
to verify, perhaps numerically, that a locally hexagonal lattice
or a similar configuration is the state with minimal energy as
conjectured in Sec. \ref{s_largegraph}. At the same time, an
extension of the model to produce extended three-dimensional
spatial lattices would also be worthwhile. This could perhaps be
done, as suggested in the original graphity model
\cite{graphity1}, by setting the preferred valence to $v_0=4$.

As the quantum graphity model is based on a Hamiltonian, it is
more akin to condensed-matter physics than to other algorithmic
approaches to building a space or spacetime from spins on graphs
\cite{Requardt:1999db,Jourjine:1984jz}. However, the Hamiltonian
approach raises an intriguing question regarding the role of
temperature. We model the geometrogenesis transition as a cooling
process that suggests the presence of a reservoir at a tunable
temperature. This could be a problem if the graph is to be
interpreted as the entire Universe. The question is then whether
this external temperature is indeed a physical temperature or some
other renormalization parameter. We believe the model needs to be
understood further before this and similar questions can be
properly addressed.

Another important next step is to study the transition from the
high-energy to the low-energy phases and look for possible
observable remnants. The transition is a complicated process and
at this stage we only understand it in the limit where $g_V$ is
the only nonzero coupling. One possibility for progress in this
direction is to generalize the random graph process to the case
where the graph cycles have structure. A necessary part of this
project involves extending string-network condensation to
irregular graphs, a question that is of interest independently of
this work.

Finally, an intriguing goal for the model is to understand the
transition from the description of the system in terms of
microscopic $(j,m)$ variables to a more standard representation in
terms of matter and geometry. The way that we normally understand
matter and gravity is as two sets of degrees of freedom coupled by
a nonlinear relation given by the Einstein equations. Normally, we
can study each part separately: in the no-gravity limit we have
quantum field theory on a fixed background and with no matter we
have pure gravity. In our model, the dynamics of the $(j,m)$
variables serves both to organize the graph into a local regular
structure with symmetries and to give rise to the effective U(1)
matter. There is no fundamental split into gravity and matter.
Since the effective matter and the geometry are different
low-energy aspects of the same microscopic degrees of freedom,
matter and the geometry can only be decoupled in a certain limit.
It has been conjectured elsewhere that such a relation can give
rise to the Einstein equations \cite{OD,Llo}. It will be
interesting to investigate this possibility in the context of our
model.

\begin{acknowledgments}

We are grateful to  our colleagues at PI and the Spinoza Institute
for comments and discussions during the course of this work. This
project was partially supported by a grant from the Foundational
Questions Institute (fqxi.org) and an NSERC Discovery grant.
Research at Perimeter Institute for Theoretical Physics is
supported in part by the Government of Canada through NSERC and by
the Province of Ontario through MRI.
\end{acknowledgments}


\begin{thebibliography}{99}


\bibitem{CDT}
  J.~Ambj\o rn, J.~Jurkiewicz and R.~Loll,
  ``Quantum gravity, or the art of building spacetime,''
  arXiv:hep-th/0604212.

\bibitem{LQG}
  C.~Rovelli, Quantum Gravity, Cambridge U. Press, New York (2004);
  T.~Thiemann, ``Introduction to modern canonical quantum general
  relativity,'' arXiv:gr-qc/0110034.;
  A.~Ashtekar and J.~Lewandowski,
  ``Background independent quantum gravity: A status report,''
  Class.\ Quant.\ Grav.\  {\bf 21}, R53 (2004)
  [arXiv:gr-qc/0404018].

\bibitem{SF}
  E.~Bianchi, L.~Modesto, C.~Rovelli and S.~Speziale,
  ``Graviton propagator in loop quantum gravity,''
  Class.\ Quant.\ Grav.\  {\bf 23}, 6989 (2006)
  [arXiv:gr-qc/0604044].

\bibitem{OritiGFT}
  D.~Oriti,
  ``The group field theory approach to quantum gravity,''
  arXiv:gr-qc/0607032;
  D.~Oriti,
  ``Group field theory as the microscopic description of the quantum spacetime
  fluid: a new perspective on the continuum in quantum gravity,''
  arXiv:0710.3276 [gr-qc].

\bibitem{KalyanaRama:2006xg}
  S.~Kalyana Rama,
  ``A principle to determine the number (3+1) of large spacetime dimensions,''
  Phys.\ Lett.\  B {\bf 645}, 365 (2007)
  [arXiv:hep-th/0610071].

\bibitem{Vol}
  G.\ Volovik, {\em The Universe in a Helium Droplet}, Oxford University Press, 2003.

\bibitem{Vis}
  M.~Visser and S.~Weinfurtner,
  ``Analogue spacetimes: Toy models for 'quantum gravity'',''
  arXiv:0712.0427 [gr-qc].

\bibitem{Unr}
  W.~G.~Unruh,
  ``Experimental black hole evaporation,''
  Phys.\ Rev.\ Lett.\  {\bf 46}, 1351 (1981).

\bibitem{CalHu}
  E.~A.~Calzetta and B.~L.~Hu,
  ``Bose-Novae as Squeezing of Vacuum Fluctuations by Condensate Dynamics,''
  arXiv:cond-mat/0207289v3.

\bibitem{Ban}
  T.~Banks,
  ``TASI lectures on matrix theory,''
  arXiv:hep-th/9911068.

\bibitem{Hor}
  P.~Horava,
  ``Stability of Fermi surfaces and K-theory,''
  Phys.\ Rev.\ Lett.\  {\bf 95}, 016405 (2005)
  [arXiv:hep-th/0503006].

\bibitem{Mal}
  J.~M.~Maldacena,
  ``The large N limit of superconformal field theories and supergravity,''
  Adv.\ Theor.\ Math.\ Phys.\  {\bf 2}, 231 (1998)
  [Int.\ J.\ Theor.\ Phys.\  {\bf 38}, 1113 (1999)]
  [arXiv:hep-th/9711200].

\bibitem{GKP}
  S.~S.~Gubser, I.~R.~Klebanov and A.~M.~Polyakov,
  ``Gauge theory correlators from non-critical string theory,''
  Phys.\ Lett.\  B {\bf 428}, 105 (1998)
  [arXiv:hep-th/9802109].

\bibitem{Gid}
  S.~B.~Giddings,
  ``Black hole information, unitarity, and nonlocality,''
  Phys.\ Rev.\  D {\bf 74}, 106005 (2006)
  [arXiv:hep-th/0605196];
  S.~B.~Giddings, D.~Marolf and J.~B.~Hartle,
  ``Observables in effective gravity,''
  Phys.\ Rev.\  D {\bf 74}, 064018 (2006)
  [arXiv:hep-th/0512200].

\bibitem{graphity1}
  T.~Konopka, F.~Markopoulou and L.~Smolin,
  ``Quantum graphity,''
  arXiv:hep-th/0611197.

\bibitem{LevinWen}
  M.~Levin and X.~G.~Wen,
  ``Fermions, strings, and gauge fields in lattice spin models,''
  Phys.\ Rev.\ B {\bf 67}, 245316 (2003)
  [arXiv:cond-mat/0302460];
  M.~A.~Levin and X.~G.~Wen,
  ``String-net condensation: A physical mechanism for topological phases,''
  Phys.\ Rev.\ B {\bf 71}, 045110 (2005)
  [arXiv:cond-mat/0404617];
  M.~Levin and X.~G.~Wen,
  ``Quantum ether: Photons and electrons from a rotor model,''
  arXiv:hep-th/0507118.

\bibitem{CauSet} R.\ Sorkin, ``Causal Sets: Discrete Gravity (Notes for the
Valdivia Summer School)'', proceedings of the Valdivia Summer
School, edited by A. Gomberoff and D. Marolf,
arXiv:gr-qc/0309009.

\bibitem{Giesel:2006uj}
  K.~Giesel and T.~Thiemann,
  ``Algebraic quantum gravity (AQG). I: Conceptual setup,''
  Class.\ Quant.\ Grav.\  {\bf 24}, 2465 (2007)
  [arXiv:gr-qc/0607099].

\bibitem{QCHistories}
  F.~Markopoulou,
  ``Quantum causal histories,''
  Class.\ Quant.\ Grav.\  {\bf 17}, 2059 (2000)
  [arXiv:hep-th/9904009].

\bibitem{Llo} S.\ Lloyd,
    ``A theory of quantum gravity based on quantum computation,''
    quant-ph/0501135.

\bibitem{Sta} J.\ Stachel, in "Structural Foundations of Quantum Gravity,"
    edited by D.P. Rickles,
    S.R.D. French and J. Saatsi Oxford University Press (2006).

\bibitem{GJ}
  M.~R.~Garey and D.~S.~Johnson,
  ``Computers and Intractability: A Guide to the Theory of NP-Completeness,''
  New York: W. H. Freeman (1983)

\bibitem{Kogut}
  J.~B.~Kogut and L.~Susskind,
  ``Hamiltonian Formulation Of Wilson's Lattice Gauge Theories,''
  Phys.\ Rev.\ D {\bf 11}, 395 (1975).

\bibitem{Sewell}
  G.~L.~Sewell,
  ``Quantum Theory of Collective Phenomena,''
  Oxford University Press, 1986.

\bibitem{SimonSokal}
  B.~Simon, and A.~D.~Sokal,
  ``Rigorous Entropy-Energy Argument,''
  J. Stat. Phys. {\bf 25}, 4, 1981, p.679.

\bibitem{Wan:2005ip}
  Y.~Wan,
  ``2D Ising model with non-local links: A study of non-locality,''
  arXiv:hep-th/0512210.

\bibitem{ka}
  M.~Karo\'nski,
  ``Random graphs,''
   Handbook of Combinatorics, Vol. 1 (R. L. Graham, M. Gr\"otschel, L. Lov\'asz, eds.), pp. 351-380. Elsevier, Amsterdam, 1995.

\bibitem{Requardt:2003qy}
  M.~Requardt,
  ``Scale free small world networks and the structure of quantum  space-time,''
  arXiv:gr-qc/0308089.

\bibitem{Requardt:1999db}
  M.~Requardt,
  ``(Quantum) space-time as a statistical geometry of lumps in random
  networks,''
  Class.\ Quant.\ Grav.\  {\bf 17}, 2029 (2000)
  [arXiv:gr-qc/9912059].

\bibitem{Jourjine:1984jz}
  A.~N.~Jourjine,
  ``Dimensional Phase Transitions Coupling Of The Matter To The Cell Complex,''
  Phys.\ Rev.\  D {\bf 31}, 1443 (1985).

\bibitem{Wormald}
  N.~Wormald,
  ``Models of random regular graphs,'' Surveys in combinatorics, 1999 (Canterbury), (J.~D.~Lamb and D. A. Preece, eds.), pp.
  239-298,
  London Math. Soc. Lecture Note Ser., 267, Cambridge Univ. Press, Cambridge, 1999.

\bibitem{Terras}
  A.~Terras and H.~Stark,
  ``Zeta functions of finite graphs and coverings,''
  Advances in Mathematics 121 (1996), 124-165.

\bibitem{OD}
  O.~Dreyer,
  ``Why things fall,''
  arXiv:0710.4350 [gr-qc].

\end{thebibliography}
\end{document}